\documentclass[11pt,a4paper]{article}\usepackage[]{graphicx}\usepackage[]{color}
\makeatletter
\def\maxwidth{ %
  \ifdim\Gin@nat@width>\linewidth
    \linewidth
  \else
    \Gin@nat@width
  \fi
}
\makeatother

\definecolor{fgcolor}{rgb}{0.345, 0.345, 0.345}
\newcommand{\hlstd}[1]{\textcolor[rgb]{0.345,0.345,0.345}{#1}}%
\newcommand{\hlkwd}[1]{\textcolor[rgb]{0.737,0.353,0.396}{\textbf{#1}}}%

\usepackage{framed}
\makeatletter
\newenvironment{kframe}{%
 \def\at@end@of@kframe{}%
 \ifinner\ifhmode%
  \def\at@end@of@kframe{\end{minipage}}%
  \begin{minipage}{\columnwidth}%
 \fi\fi%
 \def\FrameCommand##1{\hskip\@totalleftmargin \hskip-\fboxsep
 \colorbox{shadecolor}{##1}\hskip-\fboxsep
     \hskip-\linewidth \hskip-\@totalleftmargin \hskip\columnwidth}%
 \MakeFramed {\advance\hsize-\width
   \@totalleftmargin\z@ \linewidth\hsize
   \@setminipage}}%
 {\par\unskip\endMakeFramed%
 \at@end@of@kframe}
\makeatother

\definecolor{shadecolor}{rgb}{.97, .97, .97}
\definecolor{messagecolor}{rgb}{0, 0, 0}
\definecolor{warningcolor}{rgb}{1, 0, 1}
\definecolor{errorcolor}{rgb}{1, 0, 0}
\newenvironment{knitrout}{}{} 

\usepackage{alltt}
\usepackage[T1]{fontenc}
\usepackage[utf8]{inputenc}
\usepackage{authblk}
\usepackage{amsfonts, amsmath, amsthm, amssymb}
\usepackage{mathtools}
\usepackage{hyperref}
\usepackage{fancyhdr}
\usepackage{lastpage}

\title{\textbf{The Italian Crisis and Producer Households Debt: a Source of Stability? \\ A Reproducible Research} \\ \vspace{4 mm} \large {\textit{Accepted at the Risk, Banking and Finance Society, University of Florence, New York University Stern Salomon Center, and  Warsaw School of Economics \textbf{International Credit Risk Management Conference 2014}, June 23,24 -- Warsaw, Poland}}}
\author[1]{Stefano Olgiati\thanks{stefano.olgiati@unibg.it}}
\author[2]{Gilberto Bronzini\thanks{creditadvisory@danovi.it}}
\author[1]{Alessandro Danovi\thanks{alessandro.danovi@unibg.it}}
\affil[1]{\small Department of Economics, Management and Quantitative Methods, University of Bergamo, Bergamo 24129, Italy}
\affil[2]{Credit Risk Department, Studio Danovi, Milano 20122, Italy}

\providecommand{\keywords}[1]{\textbf{\textit{Keywords ---}} #1}
\IfFileExists{upquote.sty}{\usepackage{upquote}}{}

\begin{document}
  \maketitle

\begin{abstract}
The European Credit Research Institute Research Report 2013 identifies Households debt "rapid increase and abrupt retrenchment" among the causes of macroeconomic instability in the European Union after 2008. In our research: i) we accessed the Bank of Italy Online Statistical Database on Customers and Risk for Producer Households (PH--SETCON 7010) and Non-Financial Corporations (NFC-- SETCON 7006) with R Sweave open access statistical software, which makes our analysis freely reproducible by other researchers; ii) we subset the European System of Accounts sector Households into the Bank of Italy sub-sectors Households and Producer Households, which are market producing entities limited to informal partnerships, de facto companies and sole proprietorships with up to five employees and iii) we tested the hypothesis of "rapid increase and abrupt retrenchment" of debt for this subset in Italy for the period 1996-2013.  We found that the number of PH has grown from 27\% in 1996 to 43\% in 2013 of all non-financial entities reporting to the Central Credit Registrar and that PH debt (bad debt) has been more stable with a lower Variation Coefficient of 10.3\% (14.2\%) versus 13.2\% (20.1\%) in NFC. We also found that the time series of the ratio of debt granted to NFC (numerator) versus PH (denominator) is best described (Multiple R\textsuperscript{2} 0.95) by the concavity of the 5\raisebox{0.1em}{th} degree coefficient (slope -1.22; 95\% CI -1.52 -- -0.91) of a 5\raisebox{0.1em}{th} order polynomial linear regression and by the convexity of the 2\raisebox{0.1em}{nd} degree coefficient (slope 4.26; 95\% CI 2.53 -- 5.99) for bad debt (Multiple R\textsuperscript{2} 0.47), with this concavity of debt and convexity of bad debt beginning with the Italian crisis in the second trimester of 2008. We reject the hypothesis ($p < 0.01$) of "rapid increase and abrupt retrenchment" of debt for the subset Producer Households during the Italian Crisis. We generate the hypothesis that this subset could represent a prospective source of stability relative to Non-Financial Corporation.
\end{abstract}


\section*{Background}
Producer Households are the smallest entrepreneurial entities in the European Union. According to the European System of Accounts (ESA 2010), Producer Households are distinguished from Non-Financial Corporations since they are limited to informal partnerships, de facto companies and sole proprietorships with up to five employees. The relevance of this subset in Italy can be appreciated by observing that, in the period 1996-2013, Producer Households have grown from 27\% to 43\% of all non-financial entities reporting to the Central Credit Registrar and the total credit used has grown by 300\% vs. 150\% in Non-Financial Corporations.\\

As far as households are concerned, in Italy researchers initially observed a phenomenon of declining \textit{savings} but not a growth of \textit{debt}. In 1990 Modigliani and Jappelli (Modigliani, 1993) and in 1995 Rossi and Visco (Rossi and Visco, 1995) explained such decline in life-cycle saving propensity in terms of an increase of social security wealth and mandatory severance payments.\\

In 2007 (Bank of Italy, 2007) and 2010 (Bank of Italy, 2010) the Bank of Italy has began to address household debt for the specific purpose of explaining the observed variations in net wealth.\\

In 2008 Barba and Pivetti (Barba and Pivetti, 2008) analysed the causes and long-term macroeconomic implications of a \textit{rising} household debt by addressing the contradiction in the coexistence of low-wages with high levels of aggregate demand in the United States in the period 1993--2005. \\

In 2012 Chmelar and the European Credit Research Institute (Chmelar, 2012) introduced the variable of high unemployment in the analysis, and estimated the potential double-dip danger for the economy of a rapid households debt deleveraging and \textit{decline} in peripheral European economies characterised by a high sovereign debt such as Italy.\\

In 2013 the European Credit Research Institute Research Report 2013 on Household Debt and the European Crisis (Chmelar, 2013) identified Households debt \textbf{"rapid increase and abrupt retrenchment"} between 2003 and 2012 among the causes of enduring macroeconomic instability in the European Union (Chmelar 2013, p.3, Figure 1 and p.16, Figure 9).\\

\section*{Research Idea and Research Question}
The research idea behind this paper originates from an intuition by Franco Modigliani and Tullio Jappelli (Modigliani and Jappelli, 1998, p.169) where they analysed Italy's savings life-cycle idiosyncrasies with respect to \textit{discretionary wealth}: they introduced the concept, not further investigated in their paper or other papers to the present knowledge of the authors, of \textbf {households business debt}.\\

Coherently with this research idea, this paper restricts its investigation to \textbf{producer households} debt and \textbf{tests the hypothesis of "rapid increase and abrupt retrenchment"} proposed by the European Credit Research Institute Research Report 2013 \textbf{for this subset in Italy for the period 1996-2013}.\\

\section*{Methods, Replicability of the Datasets and Reproducibility of the Findings}
In this research:

\begin{enumerate}

\item we subset the European System of Accounts sector \textit{Households} (ESA 2010 -- S.14) (ESA, 2010) into the Bank of Italy's Italian Financial Accounts (Bank of Italy, 2003) sub-sectors \textit{Households} and \textit{Producer Households}; 

\item we analysed the \textit{stability} of Producer Households debt in Italy in the period 1996-2013 in terms of its Variation Coefficient $VC=\sigma / \mu$ relative to Non-Financial Corporations; 

\item we added the analysis of bad debt and; 

\item we analysed the trend of the ratio between per capita Producer Households and Non-Financial Corporations debt and bad debt.

\end{enumerate}

For the purpose of replicability and reproducibility, we accessed the Bank of Italy Online Statistical Database (BIP ON-LINE) Information on Customers and Risk - Default Rates for Loan Facilities and Borrowers for Producer Households (PH--SETCON 7010) and Non-Financial Corporations (NFC--SETCON 7006) with R Sweave (version 3.0.2) open access statistical software, which makes both our analysis and findings freely reproducible by other researchers.

\begin{itemize}
\item Raw data have been downloaded from URI:\\ \url{http://bip.bancaditalia.it/} -- Information on Customers and Risk -- Default Rates for Loan Facilities and Borrowers -- Choose option: 2. Data: whole table (*) -- TDB30486\_ENG\_ALL

\item The raw TDB30486\_ENG\_ALL comprises 11,196 rows and 5 column variables :
\begin{knitrout}
\definecolor{shadecolor}{rgb}{0.969, 0.969, 0.969}\color{fgcolor}\begin{kframe}
\begin{alltt}
\hlkwd{head}\hlstd{(raw.data)}
\end{alltt}
\begin{verbatim}
##   SEGMENT PHENOMENA SIZE_CLASS       DATE VALUE
## 1       1  35120163       1004 2013-03-31 0.037
## 2       1  35120163       1004 2012-06-30 0.047
## 3       1  35120163       1004 2011-09-30 0.047
## 4       1  35120163       1004 2011-06-30 0.048
## 5       1  35120163       1004 2009-12-31 0.096
## 6       1  35120163       1004 2009-09-30 0.045
\end{verbatim}
\begin{alltt}
\hlkwd{tail}\hlstd{(raw.data)}
\end{alltt}
\begin{verbatim}
##       SEGMENT PHENOMENA SIZE_CLASS       DATE  VALUE
## 11191    7015 351122141       9904 1997-06-30 180845
## 11192    7015 351122141       9904 1997-03-31 164303
## 11193    7015 351122141       9904 1996-12-31 161447
## 11194    7015 351122141       9904 1996-09-30 162397
## 11195    7015 351122141       9904 1996-06-30 158706
## 11196    7015 351122141       9904 1996-03-31 512619
\end{verbatim}
\end{kframe}
\end{knitrout}

\item The raw TDB30486\_ENG\_ALL dataset has been subset for Producer Households (SETCON 7010) and Non-Financial Corporations (SETCON 7006) (See Appendix A.1: Variables used);

\item The dataset in .csv format (TDB30486\_EURO\_ENG.csv) and the R codes utilised for data analysis are available at the public repository GitHub \url{https://github.com/SAO65/IRMC2014/};

\item The output of the analysis and the database of the plots are available at the public repository GitHub \url{https://github.com/SAO65/IRMC2014/blob/master/ProdHouse_Output.csv}.

\end{itemize}

\section*{Findings}

We found that Producer Households debt (bad debt) has been more stable with a lower Variation Coefficient of 10.3\% (14.2\%) versus 13.2\% (20.1\%) in Non-Financial Corporations.\\

\begin{knitrout}
\definecolor{shadecolor}{rgb}{0.969, 0.969, 0.969}\color{fgcolor}\begin{kframe}
\begin{alltt}
\hlstd{variationCoefficient.debt.PH}
\end{alltt}
\begin{verbatim}
## [1] 0.1035
\end{verbatim}
\begin{alltt}
\hlstd{variationCoefficient.debt.NFC}
\end{alltt}
\begin{verbatim}
## [1] 0.133
\end{verbatim}
\begin{alltt}
\hlstd{variationCoefficient.baddebt.PH}
\end{alltt}
\begin{verbatim}
## [1] 0.1425
\end{verbatim}
\begin{alltt}
\hlstd{variationCoefficient.baddebt.NFC}
\end{alltt}
\begin{verbatim}
## [1] 0.2027
\end{verbatim}
\end{kframe}
\end{knitrout}

We also found that the time series of the ratio of debt granted to Non-Financial Corporations versus Producer Households is best described (Multiple R\textsuperscript{2} 0.95) by the \textit{concavity} of the 5\raisebox{0.1em}{th} degree coefficient (slope -1.22; 95\% CI -1.52 -- -0.91) of a 5\raisebox{0.1em}{th} order polynomial linear regression and by the \textit{convexity} of the 2\raisebox{0.1em}{nd} degree coefficient (slope 4.26; 95\% CI 2.53 -- 5.99) for bad debt (Multiple R\textsuperscript{2} 0.47), with this \textit{concavity} of debt (Figure 2 -- Appendix A.2: Tables 2.1 and 2.2) and \textit{convexity} of bad debt beginning with the Italian crisis in the second trimester of 2008 (Figure 1 -- Appendix A.2: Tables 1.1 and 1.2).

\subsection*{Figure 1: Relative stability in terms of the ratio between debt of Non-Financial Corporations and Producer Households in Italy in the period 1996-2013 -- SOURCE: Authors' elaborations on data from the Bank of Italy}
\begin{knitrout}
\definecolor{shadecolor}{rgb}{0.969, 0.969, 0.969}\color{fgcolor}
\includegraphics[width=\maxwidth]{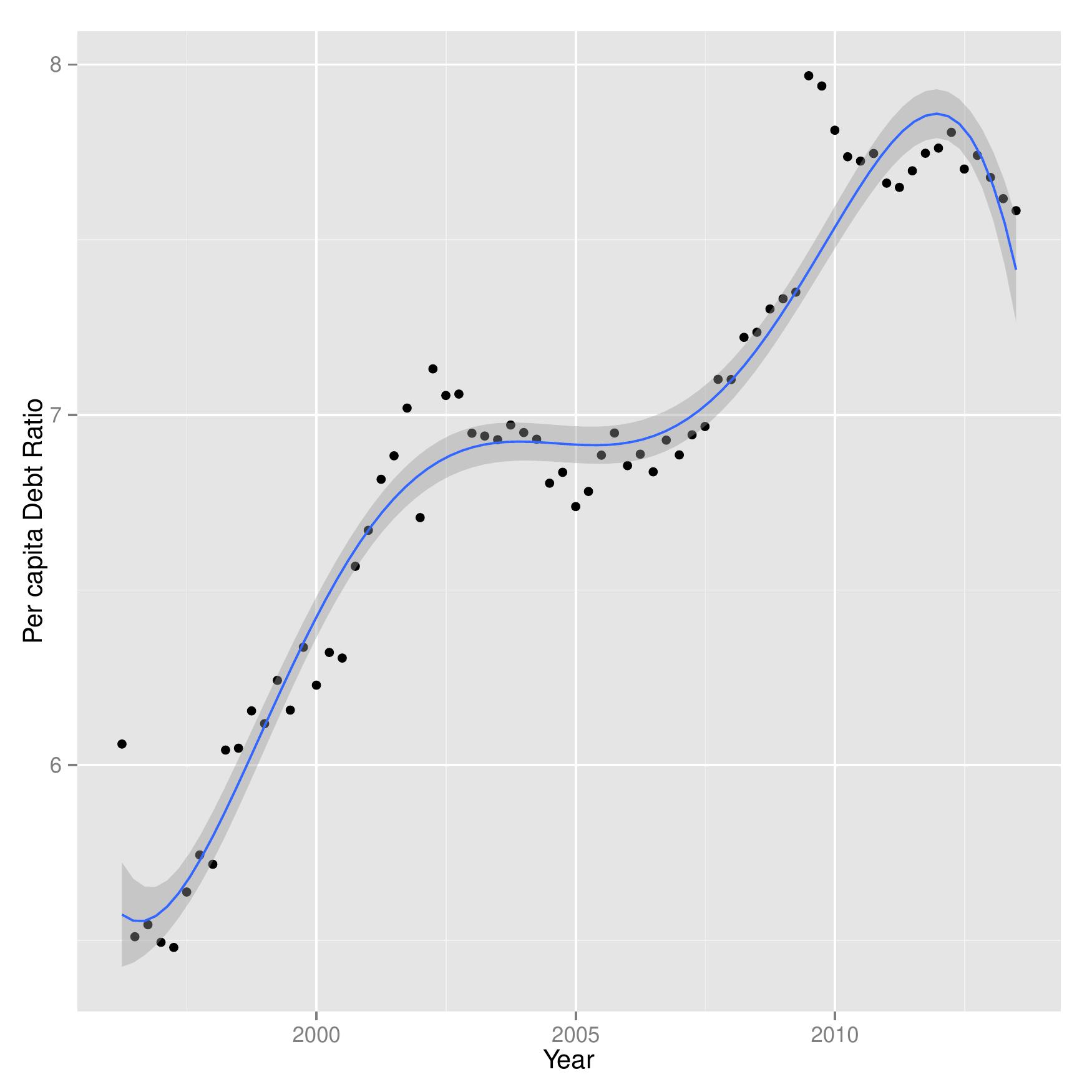} 

\end{knitrout}

\newpage
\subsection*{Figure 2: Relative stability in terms of the ratio between bad debts of Non-Financial Corporations and Producer Households in Italy in the period 1996-2013 -- SOURCE: Authors' elaborations on data from the Bank of Italy}
\begin{knitrout}
\definecolor{shadecolor}{rgb}{0.969, 0.969, 0.969}\color{fgcolor}
\includegraphics[width=\maxwidth]{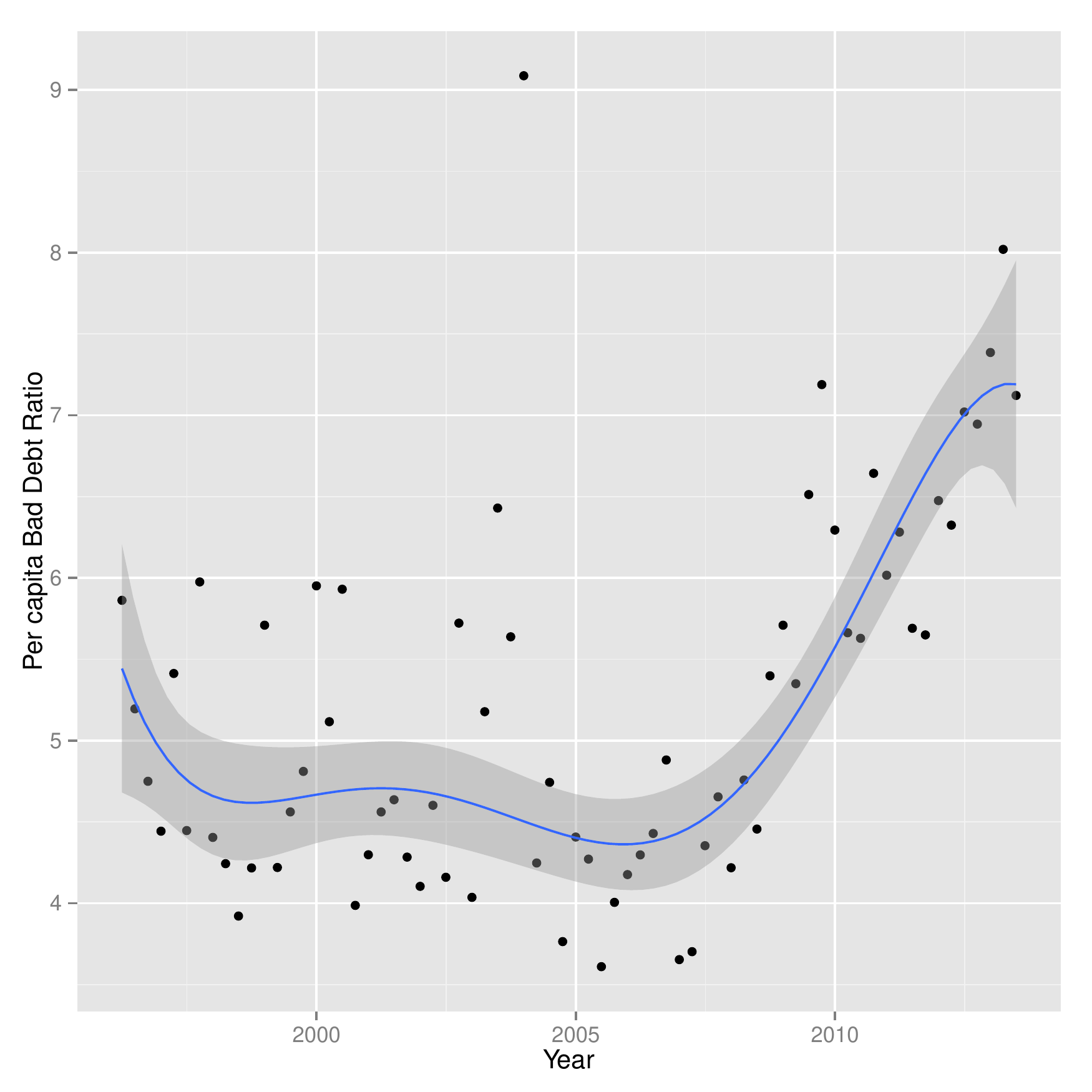} 

\end{knitrout}

\section*{Intepretation}
We reject the hypothesis ($p < 0.001$) of "rapid increase and abrupt retrenchment" of debt for the subset Producer Households in Italy in the period 1996-2013. We generate the hypothesis that Producer Households could represent a prospective source of stability relative to Non-Financial Corporations.\\



\newpage
\section*{Appendix A.1: Variables used highlighted in \textcolor{red}{red}}
\begin{itemize}

\item  \textbf{ENTE\_SEGN} -- \textcolor{red}{REPORTING INSTITUTION  3691030    BANKS, FINANCIAL COMPANIES AND OTHER INSTITUTIONS REPORTING TO THE CCR}

\item \textbf{VOCESOTVOC-- PHENOMENA OBSERVED}
\subitem \textcolor{red} {351122133   LOAN FACILITIES (EXCLUDING ADJUSTED BAD DEBTS) : CREDIT USED}  
\subitem \textcolor{red} {351122141    LOAN FACILITIES (EXCLUDING ADJUSTED BAD DEBTS) : NUMBER OF BORROWERS}  
\subitem \textcolor{red} {035120163	QUARTERLY DEFALUT RATES FOR LOAN FACILITIES : NUMER OF BORROWERS)}  
\subitem                 035120363	QUARTERLY DEFAULT RATES FOR LOAN FACILITIES : CREDIT USED  
\subitem \textcolor{red} {351121433	QUARTERLY FLOW - ADJUSTED BAD DEBTS AT T FROM PERFORMING LOANS AT T-1: AMOUNT}  
\subitem \textcolor{red} {351121441	QUARTERLY FLOW - ADJUSTED BAD DEBTS AT T FROM PERFORMING LOANS AT T-1:BORROWERS}

\item \textbf{CLASSE\_UTI --   TOTAL CREDIT USED (SIZE CLASSES)}  
\subitem                 1006	500,000 EUROS AND MORE  
\subitem                 1005	FROM >= 125,000 TO < 500,000 EUROS  
\subitem                 1004	LESS THAN 125,000 EUROS  
\subitem \textcolor{red} {9904	TOTAL (>=0)}

\item \textbf{SETCON  --  CUSTOMER SECTORS AND SEGMENTS OF ECONOMIC ACTIVITY}  
\subitem                  7015	CONSUMER HOUSEHOLDS  
\subitem                  0026	FINANCIAL COMPANIES OTHER THAN MONETARY FINANCIAL INSTITUTIONS  
\subitem                  0001	GENERAL GOVERNMENT  
\subitem \textcolor{red} {7006	NON-FINANCIAL CORPORATIONS}  
\subitem                  7009	NON-PROFIT INSTITUTIONS SERVING HOUSEHOLDS AND UNCLASSIFIABLE UNITS  
\subitem \textcolor{red} {7010	PRODUCER HOUSEHOLDS}  
\subitem                  4015	TOTAL RESIDENT SECTOR EXCLUDING MFIS
\end{itemize}

\newpage
\section*{Appendix A.2: Statistic Tables and Robustness}

\subsubsection*{Table 1.1: Confidence Intervals --- Relative stability in terms of the ratio between debt of Non-Financial Corporations and Producer Households in Italy in the period 1996-2013 -- SOURCE: Bank of Italy}
\begin{knitrout}
\definecolor{shadecolor}{rgb}{0.969, 0.969, 0.969}\color{fgcolor}\begin{kframe}
\begin{alltt}
\hlkwd{confint}\hlstd{(lm.fit.credit.ratio)}
\end{alltt}
\begin{verbatim}
##                                           2.5 %  97.5 %
## (Intercept)                               6.853  6.9255
## poly(loan.amount.producer.data$DATE, 5)1  4.903  5.5128
## poly(loan.amount.producer.data$DATE, 5)2 -1.093 -0.4834
## poly(loan.amount.producer.data$DATE, 5)3  0.229  0.8389
## poly(loan.amount.producer.data$DATE, 5)4 -0.806 -0.1961
## poly(loan.amount.producer.data$DATE, 5)5 -1.523 -0.9136
\end{verbatim}
\end{kframe}
\end{knitrout}

\subsubsection*{Table 1.2: Parameters --- Relative stability in terms of the ratio between debt of Non-Financial Corporations and Producer Households in Italy in the period 1996-2013 -- SOURCE: Bank of Italy}
\begin{knitrout}
\definecolor{shadecolor}{rgb}{0.969, 0.969, 0.969}\color{fgcolor}\begin{kframe}
\begin{alltt}
\hlkwd{summary}\hlstd{(lm.fit.credit.ratio)}
\end{alltt}
\begin{verbatim}
## 
## Call:
## lm(formula = credit.ratio ~ poly(loan.amount.producer.data$DATE, 
##     5))
## 
## Residuals:
##     Min      1Q  Median      3Q     Max 
## -0.2394 -0.1015 -0.0098  0.0508  0.5153 
## 
## Coefficients:
##                                          Estimate
## (Intercept)                                6.8890
## poly(loan.amount.producer.data$DATE, 5)1   5.2078
## poly(loan.amount.producer.data$DATE, 5)2  -0.7884
## poly(loan.amount.producer.data$DATE, 5)3   0.5339
## poly(loan.amount.producer.data$DATE, 5)4  -0.5011
## poly(loan.amount.producer.data$DATE, 5)5  -1.2185
##                                          Std. Error t value
## (Intercept)                                  0.0182  377.57
## poly(loan.amount.producer.data$DATE, 5)1     0.1527   34.12
## poly(loan.amount.producer.data$DATE, 5)2     0.1527   -5.16
## poly(loan.amount.producer.data$DATE, 5)3     0.1527    3.50
## poly(loan.amount.producer.data$DATE, 5)4     0.1527   -3.28
## poly(loan.amount.producer.data$DATE, 5)5     0.1527   -7.98
##                                          Pr(>|t|)    
## (Intercept)                               < 2e-16 ***
## poly(loan.amount.producer.data$DATE, 5)1  < 2e-16 ***
## poly(loan.amount.producer.data$DATE, 5)2  2.6e-06 ***
## poly(loan.amount.producer.data$DATE, 5)3  0.00086 ***
## poly(loan.amount.producer.data$DATE, 5)4  0.00167 ** 
## poly(loan.amount.producer.data$DATE, 5)5  3.5e-11 ***
## ---
## Signif. codes:  0 '***' 0.001 '**' 0.01 '*' 0.05 '.' 0.1 ' ' 1
## 
## Residual standard error: 0.153 on 64 degrees of freedom
## Multiple R-squared:  0.952,	Adjusted R-squared:  0.949 
## F-statistic:  255 on 5 and 64 DF,  p-value: <2e-16
\end{verbatim}
\end{kframe}
\end{knitrout}

\subsubsection*{Table 2.1: Confidence Intervals --- Relative stability in terms of the ratio between bad debts of Non-Financial Corporations and Producer Households in Italy in the period 1996-2013 -- SOURCE: Bank of Italy}
\begin{knitrout}
\definecolor{shadecolor}{rgb}{0.969, 0.969, 0.969}\color{fgcolor}\begin{kframe}
\begin{alltt}
\hlkwd{confint}\hlstd{(lm.fit.baddebt.ratio)}
\end{alltt}
\begin{verbatim}
##                                            2.5 %   97.5 %
## (Intercept)                               4.9913  5.40418
## poly(loan.amount.producer.data$DATE, 5)1  2.6247  6.07907
## poly(loan.amount.producer.data$DATE, 5)2  2.5323  5.98670
## poly(loan.amount.producer.data$DATE, 5)3 -0.1233  3.33105
## poly(loan.amount.producer.data$DATE, 5)4 -1.5252  1.92924
## poly(loan.amount.producer.data$DATE, 5)5 -3.4720 -0.01758
\end{verbatim}
\end{kframe}
\end{knitrout}

\subsubsection*{Table 2.2: Parameters --- Relative stability in terms of the ratio between bad debts of Non-Financial Corporations and Producer Households in Italy in the period 1996-2013 -- SOURCE: Bank of Italy}
\begin{knitrout}
\definecolor{shadecolor}{rgb}{0.969, 0.969, 0.969}\color{fgcolor}\begin{kframe}
\begin{alltt}
\hlkwd{summary}\hlstd{(lm.fit.baddebt.ratio)}
\end{alltt}
\begin{verbatim}
## 
## Call:
## lm(formula = baddebt.ratio ~ poly(loan.amount.producer.data$DATE, 
##     5))
## 
## Residuals:
##    Min     1Q Median     3Q    Max 
## -1.025 -0.491 -0.220  0.305  4.309 
## 
## Coefficients:
##                                          Estimate
## (Intercept)                                 5.198
## poly(loan.amount.producer.data$DATE, 5)1    4.352
## poly(loan.amount.producer.data$DATE, 5)2    4.259
## poly(loan.amount.producer.data$DATE, 5)3    1.604
## poly(loan.amount.producer.data$DATE, 5)4    0.202
## poly(loan.amount.producer.data$DATE, 5)5   -1.745
##                                          Std. Error t value
## (Intercept)                                   0.103   50.30
## poly(loan.amount.producer.data$DATE, 5)1      0.865    5.03
## poly(loan.amount.producer.data$DATE, 5)2      0.865    4.93
## poly(loan.amount.producer.data$DATE, 5)3      0.865    1.86
## poly(loan.amount.producer.data$DATE, 5)4      0.865    0.23
## poly(loan.amount.producer.data$DATE, 5)5      0.865   -2.02
##                                          Pr(>|t|)    
## (Intercept)                               < 2e-16 ***
## poly(loan.amount.producer.data$DATE, 5)1  4.2e-06 ***
## poly(loan.amount.producer.data$DATE, 5)2  6.2e-06 ***
## poly(loan.amount.producer.data$DATE, 5)3    0.068 .  
## poly(loan.amount.producer.data$DATE, 5)4    0.816    
## poly(loan.amount.producer.data$DATE, 5)5    0.048 *  
## ---
## Signif. codes:  0 '***' 0.001 '**' 0.01 '*' 0.05 '.' 0.1 ' ' 1
## 
## Residual standard error: 0.865 on 64 degrees of freedom
## Multiple R-squared:  0.472,	Adjusted R-squared:  0.431 
## F-statistic: 11.4 on 5 and 64 DF,  p-value: 6.61e-08
\end{verbatim}
\end{kframe}
\end{knitrout}

\end{document}